\documentclass[9pt,twocolumn,twoside]{osajnl}

\usepackage{amsmath}
\usepackage{graphicx}
\usepackage{amssymb}
\usepackage{upgreek}
\usepackage{algpseudocode}
\usepackage{multirow}
\usepackage{float}
\usepackage{array}

\journal{josaa} 

\setboolean{shortarticle}{false} 

\ifthenelse{\boolean{shortarticle}}{\colorlet{color2}{color2b}}{\colorlet{color2}{color2}} 

\title{Phase-error estimation and image reconstruction from digital-holography data using a Bayesian framework}

\author[1,*]{Casey~J.~Pellizzari}
\author[2]{Mark~F.~Spencer}
\author[1]{Charles~A.~Bouman}

\affil[1]{School of Electrical and Computer Engineering, Purdue University, West Lafayette, IN, 47907}
\affil[2]{Air Force Research Laboratory, Directed Energy Directorate,  KAFB, NM, 87111}

\affil[*]{Corresponding author: cpellizz@purdue.edu}

\begin{abstract}
The estimation of phase errors from digital-holography data is critical for applications such as imaging or wave-front sensing. Conventional techniques require multiple i.i.d.\ data and perform poorly in the presence of high noise or large phase errors. In this paper we propose a method to estimate isoplanatic phase errors from a single data realization. We develop a model-based iterative reconstruction algorithm which computes the maximum a posteriori estimate of the phase and the speckle-free object reflectance. Using simulated data, we show that the algorithm is robust against high noise and strong phase errors.
\end{abstract}

\begin{document}

\maketitle

\ifthenelse{\boolean{shortarticle}}{\ifthenelse{\boolean{singlecolumn}}{\abscontentformatted}{\abscontent}}{}

\section{Introduction}

Digital holography (DH) uses coherent-laser illumination and detection to gain significant improvements in sensitivity for remote-sensing applications.  Detection involves measuring the modulation of a strong reference field by a potentially-weak signal field.  This modulation allows for the detection of signals with energies equivalent to a single photon or less~\cite{winzer1998}.

In practice, DH systems are sensitive to phase errors imparted by atmospheric perturbations or flaws in the optical system.  For imaging applications, we must estimate and remove these phase errors to form a focused image~\cite{marron1993,thurman2008,thurman2008b,marron2009,marron2010,tippie2009,tippie2010,tippie2012,fienup2014}.  Similarly, for wave-front sensing applications, estimation of the phase errors represents the desired sensor output~\cite{muller1974,spencer2015,banet2016,spencer2017}.  In either case, accurately estimating the phase errors is a necessary and critical task.  

State-of-the-art techniques for estimating phase errors from DH data involve maximizing an image sharpness metric with respect to the phase errors~\cite{thurman2008,thurman2008b,marron2009,tippie2009,tippie2010,tippie2012,fienup2014}.  Typically, multiple data realizations are obtained for which the image-to-image speckle variation is decorrelated but the phase errors are identical.  For applications using pupil-plane detection, the data realizations are corrected using an initial estimate of the phase errors then inverted using a Fast Fourier Transform (FFT).  The inversion produces estimates of the complex-valued reflection coefficient, $g$, which are incoherently averaged to reduce speckle variations. The sharpness of this speckle-averaged image is then maximized to obtain the final phase-error estimate.

Image sharpening (IS) algorithms are designed to use multiple data realizations.  However, Thurman and Fienup demonstrated that in favorable conditions, the algorithms can still be used when only one realization is available~\cite{thurman2008}.   For a high signal-to-noise ratio (SNR), with several hundred detected photons per detector element, their algorithm was able to estimate atmospheric phase functions with residual root-mean-square (RMS) errors as low as $1/8~wave$.

While IS algorithms have been successfully demonstrated for estimating both isoplanatic and anisoplanatic phase errors, there remains room for improvement.  First, IS algorithms use relatively simple inversion techniques to reconstruct the magnitude squared of the reflection coefficient, $|g|^2$, rather than the real-valued reflectance, r.  The reflectance, given by $r=E[|g|^2]$, where $E[\cdot]$ indicates the expected value, is a smoother quantity with higher spatial correlation between elements in the image. We are accustomed to seeing the reflectance in conventional images and it is of greater interest for many imaging applications.  Conversely, reconstructing $|g|^2$, leads to images with high-spatial-frequency variations known as speckle. By reconstructing $r$, we can leverage its higher spatial correlation to better constrain the estimation process and potentially produce more accurate estimates of the phase errors~\cite{pellizzari2016}.

Another limitation of IS algorithms is that the process of estimating the phase errors is not tightly coupled to image reconstruction. Information about the object can be further leveraged during the phase-estimation process.  For example, this information can help rule out contributions to the phase caused by noise.  By jointly estimating both the phase errors and the reflectance, we can incorporate additional information into our estimates which helps reduce uncertainty~\cite{pellizzari2016}.

Lastly, in many practical applications it may not be possible to obtain multiple data realizations.  Particularly, in cases where there is not enough relative movement between the object and receiver to decorrelate the speckle variations between images, the phase errors are rapidly changing, or when timing requirements prohibit multiple images from being obtained.  Without multiple realizations, IS algorithms can struggle to produce accurate phase error estimates~\cite{thurman2008} and an alternate method is needed.

In this paper, we propose an improved method of DH phase recovery based on a framework of joint image reconstruction and phase-error estimation.  Our approach builds on the work of~\cite{pellizzari2016} by reconstructing the spatially-correlated reflectance, $r$, rather than the reflection coefficient, $g$, which allows for better estimates of the phase errors from less data.  We focus on reconstructing from a single data realization under isoplanatic atmospheric conditions for the off-axis pupil plane recording geometry (PPRG)~\cite{banet2016}. Our major contributions include:
\begin{enumerate}
\item We jointly compute the maximum a posteriori (MAP) estimates of the image reflectance, $r$, and phase errors, $\phi$ from a single data realization.  Joint estimation reduces uncertainty in both quantities.  Additionally, estimating $r$ rather than $g$ helps to further constrain the problem and produce images without speckle.
\item We derive the forward model for a DH system using the off-axis PPRG.  The model ensures our estimates are consistent with the physics and statistics of the remote sensing scenario.
\item We develop a technique to compute the 2D phase errors that occur in DH.  Our approach allows for the estimation of phase errors on a lower resolution grid to help reduce the number of unknowns.
\item We model the phase errors as a random variable using a Gaussian Markov Random Field (GMRF) prior model.  This approach allows us to compute the MAP estimate of the phase errors which constrains the estimate to overcome high noise and strong turbulence.
\item We compare the proposed algorithm to the image sharpening approach in~\cite{thurman2008} over a range of SNRs and atmospheric turbulence strengths.
\end{enumerate}

Overall, our experimental results using simulated, single-frame, isoplanatic data demonstrate that the proposed MBIR algorithm can reconstruct substantially higher-quality phase and image reconstructions compared to IS algorithms.

\section{Estimation Framework}
\label{Estimation_Framework}

\begin{figure}[!tb]
  \centering
\includegraphics[width=3.5in]{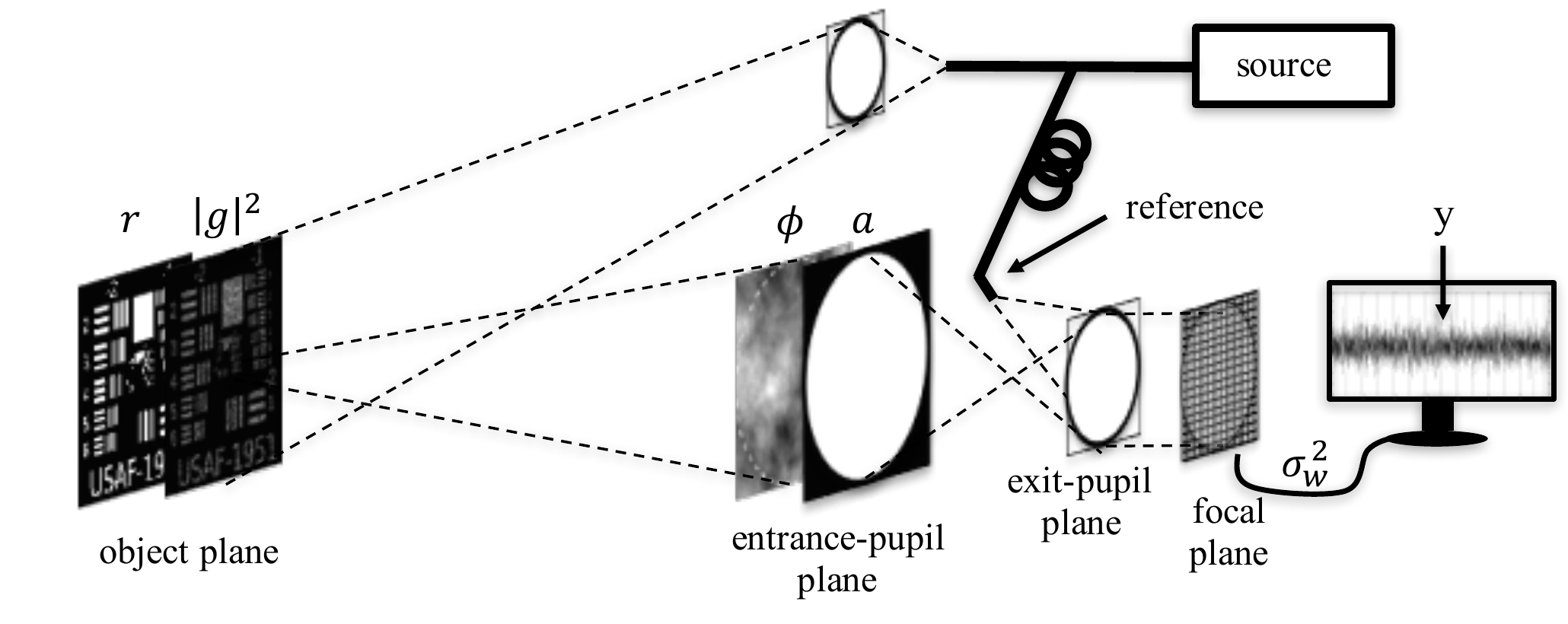}
\caption{Example DH system using an off-axis pupil plane recording geometry.  A coherent source is used to flood illuminate an object which has a reflectance function, $r$, and corresponding reflection coefficient, $g$.  The return signal is corrupted by atmospheric phase errors, $\phi$, and passes through the entrance pupil to the imaging system, $a$.  Both $a$ and $\phi$ are considered to be located in the same plane.  The entrance-pupil plane is then imaged onto a focal-plane array where it is mixed with a reference field. For simplicity, only the entrance and exit pupils of the optical system are shown. Finally, the noisy data, $y$, (with noise power $\sigma_w^2$ ) is processed to form an image and/or an estimate of $\phi$.}
\label{fig:DH_Fig}
\end{figure}

Figure~\ref{fig:DH_Fig} shows an example scenario for a DH system using the off-axis PPRG.  Our goal is to compute the MAP estimates of the reflectance, $r \in \mathbb{R}^{N}$, and the phase errors, $\phi \in \mathbb{R}^{N}$, from the noisy data, $y \in \mathbb{C}^{N}$.  Note that we are using vectorized notation for these 2D variables.  The joint estimates are given by
\begin{equation}
\begin{aligned}
	\label{eq:MAP1}
	(\hat{r},\hat{\phi}) & = \underset{(r, \phi) \in \Omega} {\mathrm{argmin}} \left \{ - \log p \left( r, \phi | y  \right) \right \}\\
	& = \underset{(r, \phi) \in \Omega} {\mathrm{argmin}} \left \{  -\log p \left( y |r,\phi \right) - \log p \left( r \right) - \log p \left( \phi \right)\right \},\\
	\end{aligned}
\end{equation}
where $\Omega$ represents the jointly-feasible set and the quantities $r$ and $\phi$ are assumed to be independent.   

To evaluate the cost function for Eq.~(\ref{eq:MAP1}), we must derive the model for $p(y|r,\phi)$.  The reader is referred to App. A for a more detailed derivation.  In short, we can represent the data using an additive noise model given by
\begin{equation}
\begin{aligned}
	\label{eq:Data}
	y = & A f+w ,
\end{aligned}
\end{equation}
where $w \in \mathbb{C}^{N}$ is the measurement noise,  $A \in \mathbb{C}^{N\times N}$ is the linear forward model operator, and $f\in \mathbb{C}^{N}$ is the complex field in the object plane.  For an object with reflectance function $g \in \mathbb{C}^{N}$, $f$ is defined as 
 \begin{equation}
	\label{eq:f}
	f = \Gamma g ,
\end{equation}
where $\Gamma \in \mathbb{C}^{N\times N}$ is a diagonal matrix that applies the object-plane, quadratic phase factor from the Fresnel propagation integral~\cite{Goodman05}.  Note that $|f_i|=|g_i| $, and therefore $r_i=E[|g_i|^2]=E[|f_i|^2]$ for all $i$. By representing the data as a function of $f$, rather than $g$, we avoid having to explicitly define $\Gamma$ in our model.  

In Eq.~(\ref{eq:Data}), the matrix $A$ accounts for the propagation and measurement geometry.  If we ignore the blurring effects caused by sampling the signal with finite-sized pixels, $A$ can be decomposed as 
 \begin{equation}
	\label{eq:A}
	A =  \mathcal{D} \left( a \right) \mathcal{D}\left( \exp \left\{ j \phi \right\} \right) D \ .
\end{equation}
Here $\mathcal{D} \left( \cdot \right)$ denotes an operator that produces a diagonal matrix from its vector argument, $a \in \mathbb{R}^{N}$ is the entrance-pupil transmission function, and $\phi$ is the phase error function previously defined.  For our purposes, $a$ is a binary circular function defining the entrance pupil of the imaging system, as shown in Fig~\ref{fig:DH_Fig}.  Finally, we choose the reconstruction parameters such that $D \in \mathbb{C}^{N \times N}$ is a two-dimensional discrete Fourier transform (DFT) matrix scaled so that $D^HD=I$.

For surfaces which are rough relative to the illumination wavelength, the vector $g$ can be modeled as a conditionally complex Gaussian random variable.  Given the underlying reflectance, $r$, of the scene, the conditional distribution is given by
\begin{equation}
	\label{eq:Dist_g}
	p(g|r) \sim CN(0, \mathcal{D}(r) ),
\end{equation}
where for any random vector $z\in  \mathbb{C}^{M}$, $CN(\mu,C)$ is the multivariate complex normal distribution with mean, $\mu$, and covariance matrix, $\Sigma$, such that
\begin{equation}
	\label{eq:CN}
	CN(\mu,C) = \frac{1}{\pi^M |\Sigma|} \exp \left\{  -(z-\mu)^H \Sigma^{-1} (z-\mu)   \right\} \ .
\end{equation}
Here, superscript $H$ indicates the Hermitian transpose. The vector, $f$, then has an identical distribution given by
\begin{equation}
\begin{aligned}
	\label{eq:Dist_f}
	p(f|r) & \sim CN(0, \Gamma \mathcal{D}(r) \Gamma^H), \\
	& = CN(0,  \mathcal{D}(r) )\ .
\end{aligned}
\end{equation}
The equality in Eq.~(\ref{eq:Dist_f}) results from the commutative property of diagonal matrices and from the fact that $\Gamma \Gamma^H=I$.

It is common for coherent detection systems to use a reference beam which is much stronger than the return signal.  For such cases, shot noise, driven by the power of the reference beam, is the dominate source of measurement noise and can be modeled as additive, zero-mean, complex Gaussian white noise~\cite{Lucke02,protopopov2009}.  The distribution of $w$ is therefore given by 
\begin{equation}
	\label{eq:Dist_w}
	p(w) \sim CN(0, \sigma_w^2 I),
\end{equation}
where $\sigma_w^2$ is the noise variance.

From Eqs.~\ref{eq:Data}-\ref{eq:Dist_w}, the likelihood function of the data, given the reflectance and the phase errors, is distributed according to~\cite{Kay93}
\begin{equation}
	\label{eq:Dist_y}
	p(y|r,\phi) \sim CN(0, A \mathcal{D}(r) A^{H}+\sigma_w^2 I) \ .
\end{equation}
Therefore, the MAP cost function is given by
\begin{equation}
\begin{aligned}
	\label{eq:MAPCost0}
c(r,\phi) = & -\log p\left( y |r,\phi \right) - \log p \left( r \right) - \log p \left( \phi \right), \\
= & ~\log  | A \mathcal{D}(r) A^{H}+\sigma_w^2 I | + y^{H} \left( A \mathcal{D}(r) A^{H}+\sigma_w^2 I\right)^{-1} y \\
  & ~~~- \log p \left( r \right) - \log p \left( \phi \right) \ .
\end{aligned}
\end{equation}
Unfortunately, the determinate and inverse make direct optimization of Eq.~(\ref{eq:MAPCost0}) an extremely computationally expensive task.

Equation~(\ref{eq:MAPCost0}) is similar to the MAP cost function in~\cite{pellizzari2016} except that the forward model has changed and we include the distribution of $\phi$ in the cost function.  Despite these differences, we use a similar optimization approach which leverages the expectation maximization (EM) algorithm and allows us to replace Eq.~(\ref{eq:MAPCost0}) with a more-tractable surrogate function.  

In order to use the EM algorithm, we need to first introduce the concept of surrogate functions.  For any function $c(x)$, we define a surrogate function, $Q(x;x')$, to be an upper-bounding function, such that $c(x) \le Q(x;x') + \kappa$, where $x'$ is the current value of $x$ which determines the functional form of $Q$ and $\kappa$ is a constant that ensures the two functions are equal at $x'$.  Surrogate functions have the property that minimization of $Q$ implies minimization of $f$.  That is 
\begin{equation}
\begin{aligned}
	\label{eq:Surrogate}
	 \left\{ Q(x;x') < Q(x';x') \right\} \Rightarrow \left\{ c(x) < c(x') \right\} \ .
	 \end{aligned}
\end{equation}

Surrogate functions are useful because in some cases it is possible to find a simple surrogate function, $Q$, that we can iteratively minimize in place of a more difficult to compute function, $c$. In fact, the EM algorithm provides a formal framework for constructing such a surrogate function. More specifically, for our problem the surrogate function of the EM algorithm takes the following form:  
\begin{equation}
\begin{aligned}
	\label{eq:Surrogate1}
	Q(r,\phi;r',\phi')  = & - \text{E} \left[ \log p(y,f~ | r,\phi) | Y =y, r', \phi' \right] \\
	& ~~~- \log p \left( r \right) - \log p \left( \phi \right),\\ 
\end{aligned}
\end{equation}
where $r'$ and $\phi'$ are the current estimates of $r$ and $\phi$, respectively.  Here we choose $f$ to be the missing data.  Evaluation of the expectation in Eq.~(\ref{eq:Surrogate1}) constitutes the E-step of the EM algorithm.  Fig.~\ref{EMAlgorithm0} shows the alternating minimization approach used for implementing the M-step of the EM algorithm.   The proposed algorithm shares the same convergence and stability properties as the algorithm in~\cite{pellizzari2016}.  More specifically, the EM algorithm converges in a stable manner to a local minima. 

Since the cost function is nonconvex, the result of the EM iterations will depend on the initial conditions and may not be the global minimum.  While it is not generally possible to compute a guaranteed global minima of the cost function in Section 4.C, we present a multi-start algorithm that we empirically found to robustly compute good local minima. 

\begin{figure}
\begin{algorithmic}
\State \bf{Repeat\{} 
\State ~~~~~~~~~~~$\hat{r} \gets \underset{r} {\mathrm{argmin}}~ Q(r,\phi';r',\phi')  $
\State ~~~~~~~~~~~$\hat{\phi} \gets \underset{\phi} {\mathrm{argmin}}~ Q(r',\phi;r',\phi')  $
\State ~~~~~~~~~~~$r' \gets  \hat{r}$
\State ~~~~~~~~~~~$\phi' \gets  \hat{\phi}$
\State \bf{\}}
\end{algorithmic}
\caption{EM algorithm for joint optimization of the MAP cost surrogate function}
\label{EMAlgorithm0}
\end{figure}

\section{Prior Models}
\label{Prior_Models}
In this section, we present the distributions used to model both the reflectance, $r$, and the phase errors, $\phi$.  In both cases, we choose to use Markov random fields (MRFs) with the form
\begin{equation}
\label{eq:gibbs}
p \left( r \right) = \frac{1}{z} \exp \left\{ - \sum_{\left\{ i,j \right\} \in \mathcal{P} } b_{i,j} \rho \left( \Delta\right) \right\},
\end{equation}
where $z$ is the partition function, $b_{i,j}$ is the weight between neighboring pixel pairs ($r_i$ and $r_j$, or $\phi_i$ and $\phi_j$), $\Delta$ is the difference between those pixel values, $\mathcal{P}$ is the set of all pair-wise cliques falling within the same neighborhood, and $\rho(\cdot)$ is the potential function~\cite{BoumanBook}.  We obtain different models for the reflectance and phase errors by choosing different potential functions.

\subsection{Reflectance}
For the reflectance function, $r$, we used a Q-Generalized Gaussian Markov Random Field (QGGMRF) potential function with the form 
\begin{equation}
\label{eq:QGGMRF}
\rho_r\left( \frac{\Delta_r}{\sigma_r}\right) = \frac{|\Delta_r|^p}{p\sigma_r^p} \left( \frac{\vert \frac{\Delta_r}{T\sigma_r} \vert^{q-p}}{1+\vert \frac{\Delta_r}{T\sigma_r} \vert^{q-p}} \right),
\end{equation}
where $T$ is a unitless threshold value which controls the transition of the potential function from having the exponent $q$ to having the exponent $p$~\cite{Thibault}.  The variable, $\sigma_r$, controls the variation in $\hat{r}$.   The parameters of the QGGMRF potential function affect its shape, and therefore the influence neighboring pixels have on one another.

While more sophisticated priors exist for coherent imaging~\cite{pellizzari2017}, the use of a QGGMRF prior is relatively simple and computationally inexpensive, yet effective~\cite{pellizzari2016}.  The low computational burden is important for DH applications where we desire near-real-time, phase-error estimation, such as with wave-front sensing.  

\subsection{Phase Errors}

In practice, the phase-error function may not change much from sample to sample, or we may wish to correct the phase at a defined real-time frame rate using a deformable mirror with fewer actuators than sensor measurements.  Therefore, to reduce the number of unknowns, we allow the phase-error function, $\phi$, to be modeled on a grid that has lower resolution than the measured data.  We denote the subsampled function as $\bar{\phi}$. 

Using $\bar{\phi}$ allows us to estimate only a single phase value for multiple data samples. This technique reduces the number of unknowns when estimating $r$ and $\phi$ from $2N$ to $N(1+1/n_b^2)$, where $n_b$ is the factor of subsampling used in both dimensions.  To scale $\bar{\phi}$ to the resolution of $\phi$ for use in the forward model, we use a nearest-neighbor-interpolation scheme given by
\begin{equation}
	\label{eq:phiphi}
	\phi = F \bar{\phi} \ .
\end{equation}
Here, $F$ is an $N \times N/n_b^2$ interpolation matrix with elements in the set $[0,1]$.

To model the distribution of unwrapped phase error values in $\bar{\phi}$, we use a Gaussian potential function given by
\begin{equation}
\label{eq:GMRF}
\rho_{\bar{\phi}}\left( \frac{\Delta_{\bar{\phi}}}{\sigma_{\bar{\phi}}}\right) = \frac{|\Delta_{\bar{\phi}}|^2}{2\sigma_{\bar{\phi}}^2} ,
\end{equation}
where $\sigma_{\bar{\phi}}$ controls the variation in $\hat{\bar{\phi}}$.  The Gaussian potential function enforces a strong influence between neighboring samples which generates a more-smoothly varying output when compared to QGGMRF.  Since the unwrapped phase function will not have edges or other sharp discontinuities, the GMRF is more appropriate than QGGMRF in this case.  An added benefit is that it is less computationally expensive than the QGGMRF model which allows for fast evaluation.

\section{Algorithm}
\label{algorithm}
In this section, we present the details for how we execute the proposed EM algorithm shown in Fig.~\ref{EMAlgorithm0}.  The discussion describes how we evaluate the surrogate function in the E-step and how we optimize it in the M-step.  

\subsection{E-step - Surrogate Function Evaluation }
\label{estep}
Using the models given by Eqs.~(\ref{eq:Dist_f}-\ref{eq:Dist_y}) and (\ref{eq:gibbs}-\ref{eq:GMRF}), along with the conditional posterior distribution of $f$ given $y$ and $r$, we can evaluate the surrogate function of Eq.~(\ref{eq:Surrogate1}).  Its final form is given by 
\begin{equation}
\begin{aligned}
	\label{eq:MAPCost4b}
	Q(r,\phi;r',\phi') = &  - \frac{1}{\sigma_w^2} 2 \text{Re} \left\{ y^HA_{\phi} \mu \right\}  + \log | \mathcal{D}( r)| \\
	& ~~~+ \sum_{i=1}^{N} \frac{1}{r_i}~\left( C_{i,i} + | \mu_i|^2\right)  + \sum_{\left\{ i,j \right\} \in \mathcal{P} } b_{i,j} \rho_r \left( \frac{\Delta_r}{\sigma_r}\right)\\
	& ~~~ + \sum_{\left\{ i,j \right\} \in \mathcal{P} } b_{i,j} \rho_{\bar{\phi}} \left( \frac{\Delta_{\bar{\phi}}}{\sigma_{\bar{\phi}}}\right) \ .
	 \end{aligned}
\end{equation}
Here, $A_{\phi}$ indicates that the matrix $A$ is dependent on $\phi$, $\mu_i$ is the $i^{th}$ element of the posterior mean given by 
\begin{equation}
	\label{eq:mu_g}
	\mu=C  \frac{1}{\sigma_w^2 } A_{\phi'}^H y,
\end{equation}
and $C_{i,i}$ is the $i^{th}$ diagonal element of the posterior covariance matrix given by 
\begin{equation}
\begin{aligned}
	\label{eq:C_g}
	C & = \left[\frac{1}{\sigma_w^2 } A_{\phi'}^HA_{\phi'}+  \mathcal{D}( r)^{-1}\right]^{-1} \ .
\end{aligned}
\end{equation}

We simplify the evaluation of Eq.~(\ref{eq:MAPCost4b}) by approximating the posterior covariance as
\begin{equation}
\begin{aligned}
	\label{eq:C_g2}
	C  \approx \mathcal{D} \left( \frac{\sigma_w^2}{1+\frac{\sigma_w^2} {r} }\right) ,
\end{aligned}
\end{equation}
which requires that
\begin{equation}
	\label{eq:AHA}
	A_{\phi}^HA_{\phi} = D^H \Phi^H \Lambda^H \Lambda \Phi D \approx I \ .
\end{equation}
For a square entrance-pupil function filling the detector's field of view, $\Lambda = I$ and the approximations of Eqs.~(\ref{eq:C_g2}) and ~(\ref{eq:AHA}) become exact as both $D$ and $\Phi$ are unitary matrices.  When $\Lambda$ represents a circular aperture, then the relationship of Eq.~(\ref{eq:AHA}) is approximate. However, since it dramatically simplifies computations, we will use this approximation and the resulting approximation of Eq.~(\ref{eq:C_g2}) in all our simulations when computing the $Q$ function for the EM update. In practice, we have found this to work well.

\subsection{M-step - Optimization of the Surrogate Function}

The goal of the M-step is to minimize the surrogate function according to
\begin{equation}
\begin{aligned}
	\label{eq:MStep}
	(\hat{r},\hat{\phi}) & = \underset{(r, \phi)} {\mathrm{argmin}} \left \{ Q(r,\phi;r',\phi') \right \} \ .
	\end{aligned}
\end{equation}
The alternating optimization shown in Fig.~\ref{EMAlgorithm0} is used to minimize $Q$ with respect to $r$ and $\phi$ during each iteration of the EM algorithm.  

We use Iterative Coordinate Descent (ICD) to update the reflectance~\cite{BoumanBook}.  In particular, ICD works by sequentially minimizing the entire cost function with respect to a single pixel, $r_s$.  By considering just the terms in Eq.~(\ref{eq:MAPCost4}) which depend on a pixel $r_s$, the cost function is given by
\begin{equation}
\begin{aligned}
	\label{eq:MAP_8}
	q_s(r_s;r',\phi')  =   \log r_s +  \frac{C_{s,s}+| \mu_s|^2}{ r_s} + \sum_{ j  \in \partial s } b_{s,j} \rho_r \left( \frac{r_s-r_j}{\sigma_r}\right)  ,
 \end{aligned}
\end{equation}
where the sum over $j  \in \partial s $ indicates a sum over all neighbors of the pixel $r_s$.  We carry out the minimization of Eq.~(\ref{eq:MAP_8}) with a 1D line search over $\mathbb{R}^+$.

To minimize Eq.~(\ref{eq:MAPCost4b}) with respect to the phase-errors, we consider just the terms which depend on $\bar{\phi}$.  The phase-error cost function becomes
\begin{equation}
\begin{aligned}
	\label{eq:phi_cost1}
	q(\phi;r',\phi') & =   - \frac{1}{\sigma_w^2} 2 \text{Re} \left\{ y^HA_{\phi} \mu \right\}  + \frac{1}{2\sigma_{\bar{\phi}}^2} \sum_{\left\{ i,j \right\} \in \mathcal{P} } b_{i,j} |\Delta_{\bar{\phi}}|^2  . \
\end{aligned}
\end{equation}
We use ICD to sequentially minimize the cost with respect to each element of the subsampled phase-error function, $\bar{\phi}$.  For element $\bar{\phi}_p$, corresponding to a $n_b\times n_b$ group of data samples with indices in the set $B$, the cost function becomes 
\begin{equation}
\begin{aligned}
	\label{eq:phi_cost2}
	q(\bar{\phi}_p;r',\phi') & =   - |\chi| \cos \left( \phi_{\chi} - \bar{\phi}_p \right)  + \frac{1}{2\sigma_{\bar{\phi}}^2} \sum_{j \in \partial p} b_{i,j} |\bar{\phi}_p-\bar{\phi}_j|^2  ,
\end{aligned}
\end{equation}
where
\begin{equation}
	\label{eq:chi}
	\chi = \frac{2}{\sigma_w^2} \sum_{i=1}^{n_b^2} y_{B(i)}^* [D\mu]_{B(i)} ,
\end{equation}
and $j\in \partial p$ is an index over neighboring phase samples on the lower resolution grid.  We minimize Eq.~(\ref{eq:phi_cost2}) using a 1D line search over $\bar{\phi}_p \in [\bar{\phi}^{**}-\pi, \bar{\phi}^{**}+\pi]$, where $\bar{\phi}^{**}$ minimizes the prior term.  By minimizing Eq.~(\ref{eq:phi_cost2}), we obtain an estimate of the unwrapped phase errors.  Finally, we obtain the full-resolution estimate of the unwrapped phase errors, $\phi$, according to Eq.~(\ref{eq:phiphi}).

\subsection{Initialization and Stopping Criteria}

Figure~\ref{EMAlgorithm} summarizes the steps of the EM algorithm.  To determine when to stop the algorithm, we can use either a set number of iterations, $N_K$, or a metric such as
 \begin{equation}
	\label{eq:StopCrit}
	\epsilon = \frac{|| r^k-r^{k-1}||}{||r^{k-1}||},
\end{equation}
where $k$ is the iteration index and we stop the algorithm when $\epsilon$ falls below a threshold value of $\epsilon_{T}$.  

\begin{figure}
\noindent \rule{3.5in}{1pt}
\begin{algorithmic}
\State \textbf{EM \{}
\State {Inputs:} $y,~r',\phi',~\sigma_r,~\sigma_{\bar{\phi}},~\sigma_w^{2},~q,~p,~T,~b,~(\text{either}~N_K~\text{or}~\epsilon_T)$
\State {Outputs:} $\hat{r},\hat{\phi}$
\While{  $k<N_K$ or $\epsilon > \epsilon_T$ }
    \State $C \gets \mathcal{D} \left( \frac{\sigma_w^2}{1+\frac{\sigma_w^2} {r'} }\right)$ 
    \State $\mu \gets  C  \frac{1}{\sigma_w^2 } A_{\phi'}^H y,$
     \ForAll{$s \in S$}
      \State $r_s \gets \underset{r_s \in \mathbb{R}^+} {\mathrm{argmin}}  \left\{ q_r(r_s;r',\phi') \right\}$
    \EndFor 
     \ForAll{$p \in P$}
       \State $\phi_{p} \gets  \underset{\phi} {\mathrm{argmin}} \left\{ q_{p}(\bar{\phi}_p;r',\phi') \right\}$
     \EndFor
     \State $\phi \gets F \bar{\phi}$
     \EndWhile
     \State \textbf{\}}
\end{algorithmic}
\noindent \rule{3.5in}{1pt}
\caption{EM algorithm for the MAP estimates of $r$ and $\phi$. Here, $S$ is the set of all pixels and $P$ is the set of all phase-error elements.}
\label{EMAlgorithm}
\end{figure}

The EM algorithm in Fig.~\ref{EMAlgorithm} is initialized according to
\begin{equation}
\label{eq:Initialization2}
	\begin{aligned}
	r & \gets |A^Hy|^{\circ 2 }, \\
	\sigma_r & \gets \frac{1}{\gamma} \sqrt{ s^2\left(r \right)}, \\
	\end{aligned}
\end{equation}
where $|\cdot|^{\circ 2}$ indicates the element-wise magnitude square of a vector, $\gamma$ is a unitless parameter introduced to tune the amount of regularization in $r$, and $s^2(\cdot)$ computes the sample variance of a vector's elements~\cite{forbes2011}. 

We use a heuristic similar to that used in~\cite{pellizzari2016} to iteratively initialize the phase-error estimate.  Fig.~\ref{MBIRAlgorithm} details the steps of this iterative process.  The initial estimate of the phase-error vector is simply $\phi \gets \mathbf{0}$.  Next, for a set number of outer-loop iterations, $N_L$, we allow the EM algorithm to run for $N_K$ iterations.  At the beginning of each outer-loop iteration, we reinitialize according to Eq.~(\ref{eq:Initialization2}).

After the outer loop runs $N_L$ times, we again reinitialize according to Eq.~(\ref{eq:Initialization2}) and run the EM algorithm until it reaches the stopping threshold $\epsilon_T$.  A Gaussian prior model for $r$ works best in the outer loop for the initialization of $\phi$.  Specifically, we use $q=2$, $p=2$, $T=1$, $\gamma=2$, and $b=G(0.8)$, where $G(\sigma)$ indicates a $3\times3$ Gaussian kernel with standard deviation, $\sigma$.  Once the initialization process is complete, we can use different prior-model parameters for the actual reconstruction.   

\begin{figure}
\noindent \rule{3.5in}{1pt}
\begin{algorithmic}
\State \textbf{MBIR Algorithm \{}
\State {Inputs:} $y,~\gamma,~\sigma_w,~q,~p,~T,~b,~N_K,~N_L,~\epsilon_T$
\State {Outputs:} $\hat{r},~\hat{\phi}$ 
\State $\phi \gets \mathbf{0} $
    \For{$i=1:N_L$}
      \State $r \gets |A^Hy|^{\circ 2 }$, $\sigma_r \gets \frac{1}{\gamma} \sqrt{\text{var} \left(r \right)}$
      \State $\phi \gets$ \textbf{EM} $\left\{y,r,\phi,\sigma_r,\sigma_{\bar{\phi}},\sigma_w^2,2,2,1,G(0.8),N_K\right\}$
    \EndFor
    \State $r \gets |A^Hy|^{\circ 2 }$, $\sigma_r \gets \frac{1}{\gamma} \sqrt{\text{var} \left(r \right)}$
      \State $r,\phi \gets$ \textbf{EM} $\left\{y,r,\phi,\sigma_r,\sigma_{\bar{\phi}},\sigma_w^2,q,p,T,b,\epsilon_T\right\}$
     \State \textbf{\}}
\end{algorithmic}
\noindent \rule{3.5in}{1pt}
\caption{Algorithm which initializes and runs the EM algorithm.  An iterative process is used to initialize the phase error vector $\phi$.}
\label{MBIRAlgorithm}
\end{figure}

\section{Methods}

\begin{figure}[!tb]
  \centering
\includegraphics[width=2.5in]{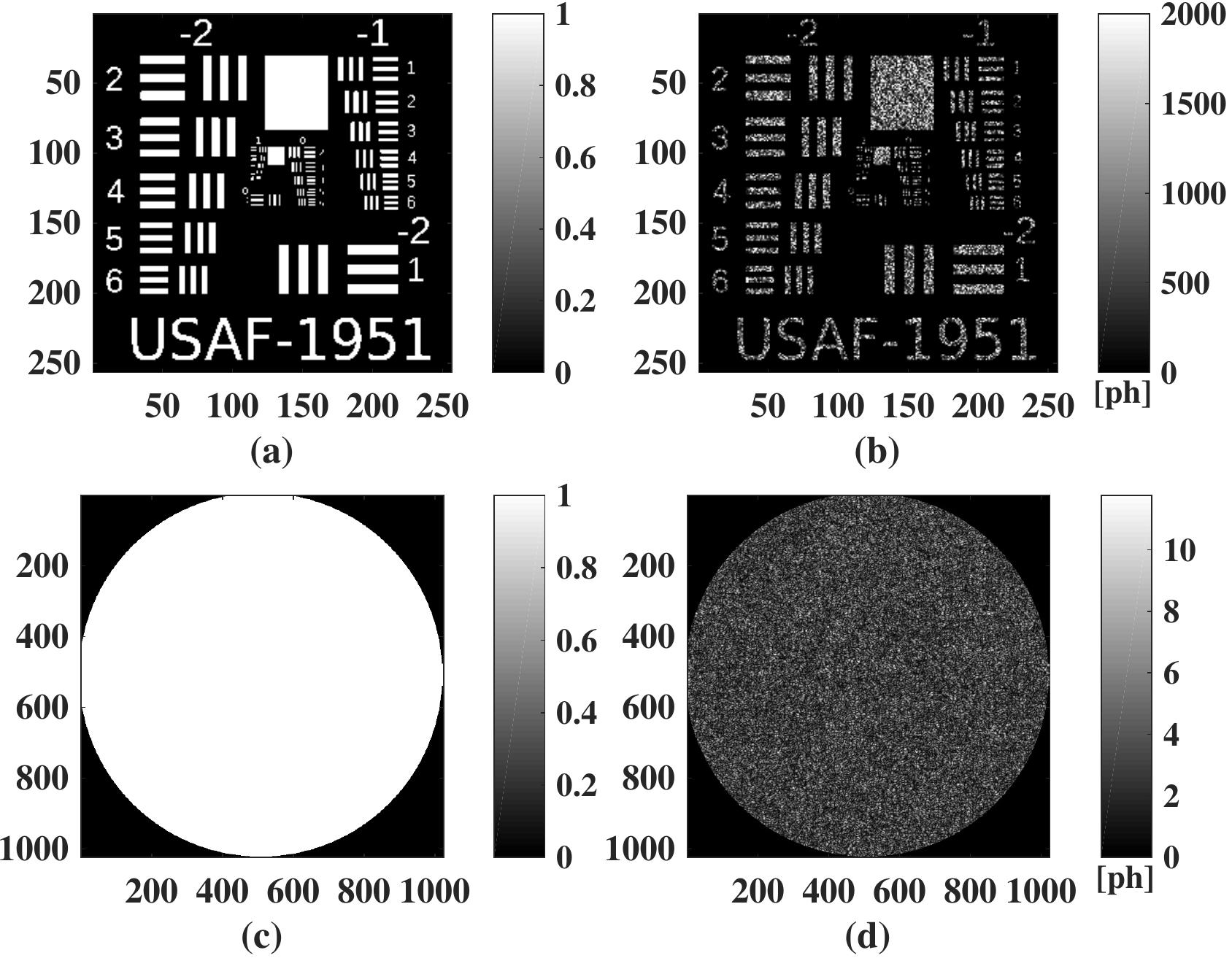}
\caption{(a) $256\times256$ reflectance function used for generating simulated data, (b) intensity of the reflected field in object plane, $|f|^{\circ 2}$, (c) entrance-pupil transmission function, $a(m,n)$, and (d) intensity of the field passing through the entrance-pupil.}
\label{fig:Sim}
\end{figure}

To compare the proposed algorithm to an IS algorithm from~\cite{thurman2008}, we generated simulated data using an approach similar to that of~\cite{thurman2008}.  Figure~\ref{fig:Sim} (a) shows the $256\times 256$ reflectance function that we multiplied by an incident power, $I_0$, to get the object intensity, $I(p,q)$.  Figure~\ref{fig:Sim} (b) shows the magnitude squared of the corresponding reflection coefficient generated according to $f \sim CN(0, \mathcal{D}(I_0r) )$.   

In accordance with the geometry shown in Fig.~\ref{fig:DH_Fig}, we used a fast Fourier transform (FFT) to propagate the field, $f$, from the object plane to the entrance-pupil plane of the DH system.  The field was padded to $1024\times 1024$ prior to propagation.  Next, we applied a phase error, $\phi(m,n)$ and the entrance pupil transmission function, $a(m,n)$, shown in  Figure~\ref{fig:Sim} (c).  Figure.~\ref{fig:Sim} (d) is an example of the intensity of the field passing through the entrance pupil which has a diameter, $D_{ap}$, equal to the grid length.

\begin{figure}[!tb]
  \centering
\includegraphics[width=2.5in]{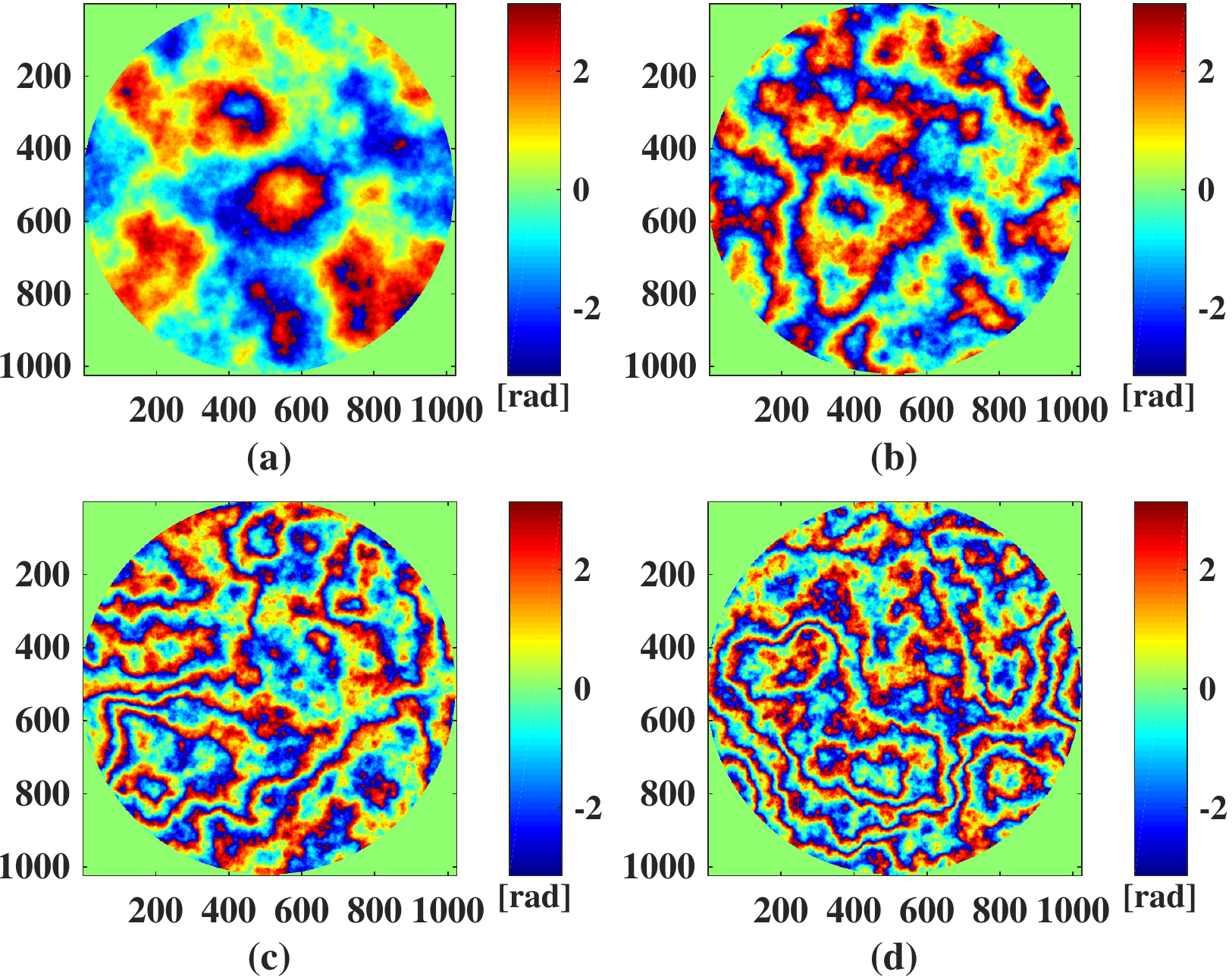}
\caption{Atmospheric phase errors for $D_{ap}/r_0$ values of (a) 10, (b) 20, (c) 30, and (d) 40.  Values shown are wrapped to $[-\pi,\pi)$}
\label{fig:PhzErrors}
\end{figure}

To generate the atmospheric phase errors, we used an FFT-based technique described in~\cite{schmidt2010}.
Using a Kolmogorov model for the refractive-index power-spectral density (PSD), we generated random coefficients for the spatial-frequency components of the atmospheric phase, then we used an inverse FFT to transform to the spatial domain.  To set the turbulence strength, we parameterized the Kolmogorov PSD by the coherence length, $r_0$, also known as the Fried parameter~\cite{andrews2005,schmidt2010}.  The value of $r_0$ specifies the degree to which the phase of a plane-wave passing through the turbulent medium is correlated. Two points in the phase function which are separated by a distance greater than $r_0$ will typically be uncorrelated.  In this paper, we report turbulence strength using the ratio $D_{ap}/r_0$ which is related to the degrees of freedom in the atmospheric phase.  We simulated data for $D_{ap}/r_0$ values of $10,~20,~30,$ and $40$.  Figure~\ref{fig:PhzErrors} shows examples of the wrapped phase errors for each case.

After we added phase errors to the propagated field and applied the aperture function shown in Fig.~\ref{fig:Sim}, we mixed the signal with a modulated reference beam and detected the resultant power.  Following~\cite{thurman2008}, the reference-beam power was set at approximately 80\% of the well depth per detector element, i.e., 80\% of $5\times 10^4 ~photoelectron~(pe)$.  We modeled Gaussian read noise with a standard deviation of $40~pe$ and digitized the output to $12~bits$.  After detection, we demodulated the signal to remove the spatial-frequency offset from the reference beam, low-pass filtered to isolate the signal of interest, and decimated to obtain a $256\times 256$ data array.\footnote{It is typical for this process to be carried out by taking an FFT, windowing a small region around the desired signal spectrum, and taking an inverse FFT.}  The resulting data was represented by Eq.~(\ref{eq:Data}) after vectorization.  

We generated data over a range of SNRs which we define as
\begin{equation}
\label{eq:SNR}
	\text{SNR} = \frac{s^2(Af)}{s^2(w)} ,
\end{equation}
where $s^2(\cdot)$ is the sample-variance operator used in Eq.~(\ref{eq:Initialization2}).  For optically-coherent systems, SNR is well approximated by the average number of detected signal photons per pixel~\cite{Lucke02,haus2012}.  At each turbulence strength, and at each SNR, we generated 18 i.i.d.\ realizations of the data.  We then processed each i.i.d.\ realization independently and computed the average performance over the 18 independent cases. 

To measure the distortion between the reconstructed images, $\hat{r}$, and the simulation input, $r$, we used normalized root mean square error (NRMSE) given by
 \begin{equation}
	\label{eq:NRMS}
	\text{NRMSE} = \sqrt{\frac{|| \alpha^* \hat{r}-  r ||^2}{|| r ||^2}} ,
\end{equation}
where
 \begin{equation}
	\label{eq:alpha}
	\alpha^* = \underset{\alpha} {\mathrm{argmin}} \left\{  || \alpha \hat{r}-  r ||^2\right\} ,
\end{equation}
is the least-squares fit for any multiplicative offset between $r$ and $\hat{r}$. 


To measure distortion between the reconstructed phase error, $\hat{\phi}$, and the actual phase error, $\phi$, we calculated the Strehl ratio according to
 \begin{equation}
	\label{eq:strehl}
	S = \frac{ \left[ |\text{FFT} \left\{ a(m,n) e^{j [\hat{\phi}(m,n)-\phi(m,n)]}\right\} |^2 \right]_{\text{max}}}{\left[ |\text{FFT} \left\{ a(m,n) \right\} |^2  \right] _{\text{max}}},
\end{equation}
where $\text{FFT} \left\{ \cdot \right\}$ is the FFT operator and $\left[ \cdot \right]_{\text{max}}$ indicates that we take the maximum value of the argument.  The function, $a(m,n)$, is a binary function that represents the aperture in the observation plane.   It takes on the value $1$ inside the white dotted circle shown in Fig.~\ref{fig:Sim} and $0$ outside.  

Digital phase correction of DH data is analogous to using a piston-only, segmented deformable mirror (DM) in adaptive optics.  Therefore, using the DM fitting error from~\cite{hardy1998} and Mar\'{e}chal's approximation, we can compute the theoretical limit of the Strehl ratio as
 \begin{equation}
	\label{eq:strehl_lim}
	S_{max} = e^{- 1.26 \left( \frac{d}{r_0} \right)^{\frac{5}{3}}} ,
\end{equation}
where $d$ is the spacing between correction points in $\hat{\phi}$.

Using NRMSE and Strehl ratio, we compared performance of the proposed algorithm to the IS approach presented in~\cite{thurman2008} using a point-by-point estimate of $\phi$ and the $M_2$ sharpness metric.   The algorithm computes the phase-error estimate according to
 \begin{equation}
	\label{eq:phz_est_is}
	\hat{\phi} = \underset{\phi} {\mathrm{argmax}} \left \{ - \sum_{p,q} \left( | D^H \mathcal{D}(\exp \left\{ j \phi \right\})^H y|^{\circ 2}\right)^{\circ 0.5} \right \},
\end{equation}
where  $^{\circ}$ indicates the application of an exponent to each element.  Following the process described in~\cite{thurman2008}, we used 20 iterations of conjugate gradient to optimize Eq.~(\ref{eq:phz_est_is}) and the algorithm was initialized with a $15^{th}$ order Zernike polynomial found through an iterative process.

For the proposed MBIR algorithm, we allowed the outer initialization loop to run $N_L=2\times10^2$ times, with $N_K=10$ EM iterations each time.  We kept $N_L$ constant for all reconstructions.  Once the iterative initialization process was complete, we set a stopping criteria of $\epsilon_T=1\times10^{-4}$ and let the EM algorithm run to completion.  We used $q=2$, $p=1.1$, $T=0.1$, $\gamma=2$, and $b=G(0.1)$ as the QGGMRF prior parameters for image reconstruction.   Additionally, we used $n_b=2$, $\sigma_{\bar{\phi}}=0.1$ and $b=G(0.1)$ for the phase error prior parameters.  Using $n_b=2$ gives a total number of unknowns of $5/4 N$.  While varying $\sigma_{\bar{\phi}}$ with turbulence strength may produce more optimal results, we kept it fixed for simplicity.  We found these parameters to work well over a wide range of conditions.

\section{Experimental Results}

\begin{figure*}[!tb]
  \centering
\includegraphics[width=7.5in]{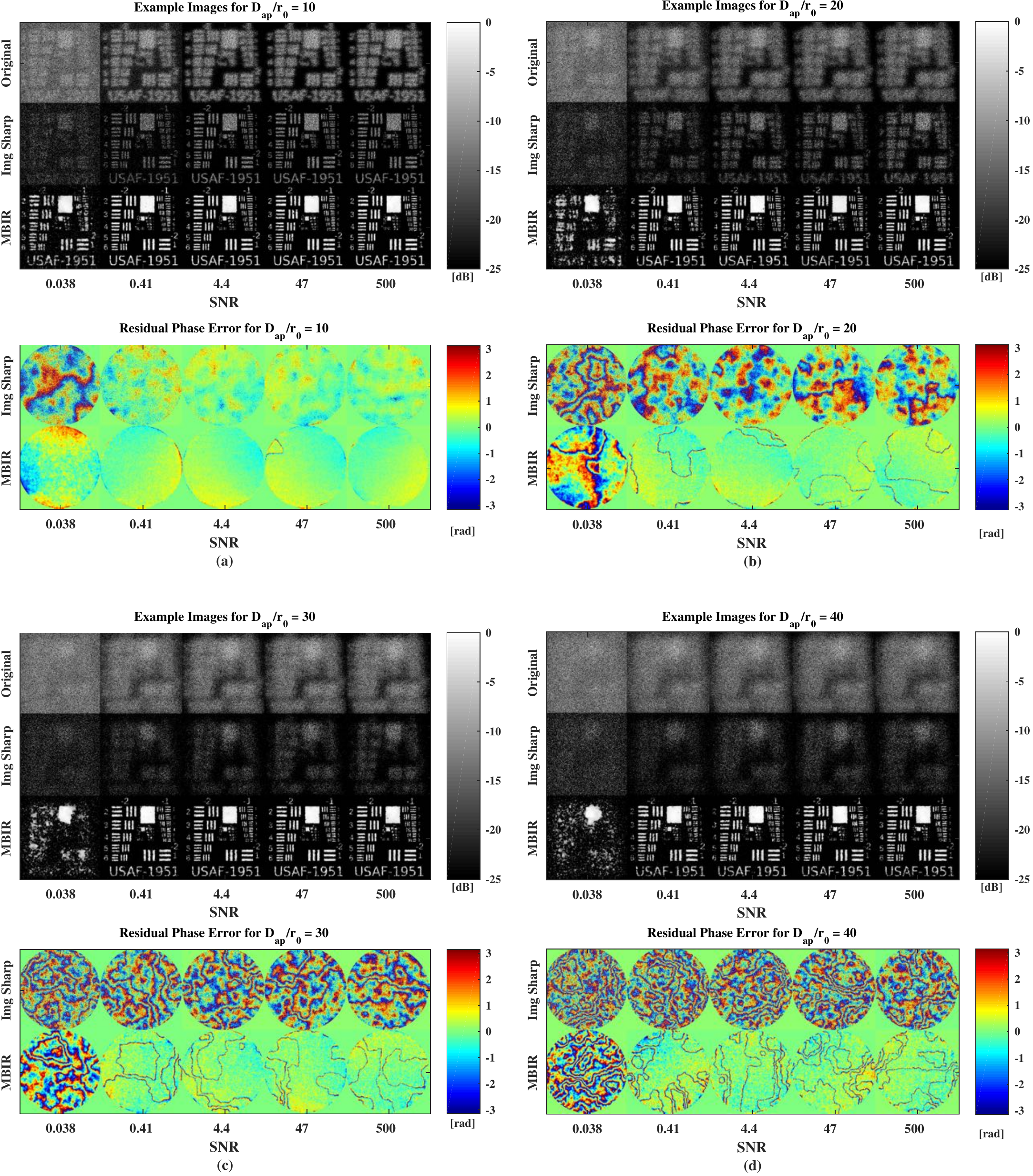}
\caption{Example images and residual phase errors for $D_{ap}/r_0$ of 10 (a),  20 (b), 30 (c), and 40 (d). The top row of each image is the original blurry image.  These examples represent the reconstructions with the median Strehl ratio at the chosen SNRs.  Note that we only show five of the 20 SNR values.  The images are shown using a log-based dB scale and the residual phase errors are wrapped to $[-\pi,\pi)$.}
\label{fig:results2}
\end{figure*}

Figure~\ref{fig:results2} shows example reconstructions for a subset of the results.  Each block of images shows the reconstructions corresponding to the median Strehl ratio of the 18 i.i.d.\ data sets.  Note that we only show five of the 20 SNR levels for each turbulence strength.  The top row of each image block shows the original blurry images, the middle shows the IS reconstructions, and the bottom shows the MBIR reconstruction.  To compress the large dynamic range that occurs in coherent imagery, we present the results using a log-based dB scale given by $r_{dB}=10 \log_{10} (\tilde{r})$, where $\tilde{r}\in \left(0,1 \right]$ is the normalized reflectance function.  The residual phase errors, wrapped to $[-\pi,\pi)$, are shown below each image block.  To aid in viewing the higher-order residual phase errors in Fig.~\ref{fig:results2}, we removed the constant and linear phase  components, piston, tip, and tilt. The values were estimated according to
\begin{equation}
(\hat{p}, \hat{t}_x, \hat{t}_y) = \underset{(p, t_x, t_y)} {\mathrm{argmin}} \sum_{m,n} | \text{W} \left\{ \phi_r (m,n) - \left(p+ t_x m + t_y n \right) \right\} |^2,
\end{equation}
where $\text{W} \left\{\cdot \right\}$ is an operator that wraps its argument to $[-\pi,\pi)$,  $\phi_r$ is the residual phase error, $p$ is a constant, and $t_x,t_y$ are the linear components in the $x$ and $y$ dimensions, respectively.

The examples in Figure~\ref{fig:results2} show that the IS algorithm is able to correct most of the phase errors for  $D_{ap}/r_0=10$ and some of the errors  at $D_{ap}/r_0=20$.  For $D_{ap}/r_0=30$ and $D_{ap}/r_0=40$, the IS images are blurred beyond recognition.  In contrast, the proposed MBIR algorithm produced focused images for all but the lowest SNR reconstructions.  In addition, we see how the MBIR reconstruction has significantly less speckle variation.  The MBIR results also show that in many cases, the residual-phase errors contain large-scale patches separated by wrapping cuts.  However, as Fig.~\ref{fig:results2} shows, the patches are approximately modulo-$2\pi$ equivalent and still produce focused images.  

\begin{figure*}[!tb]
  \centering
\includegraphics[width=6.5in]{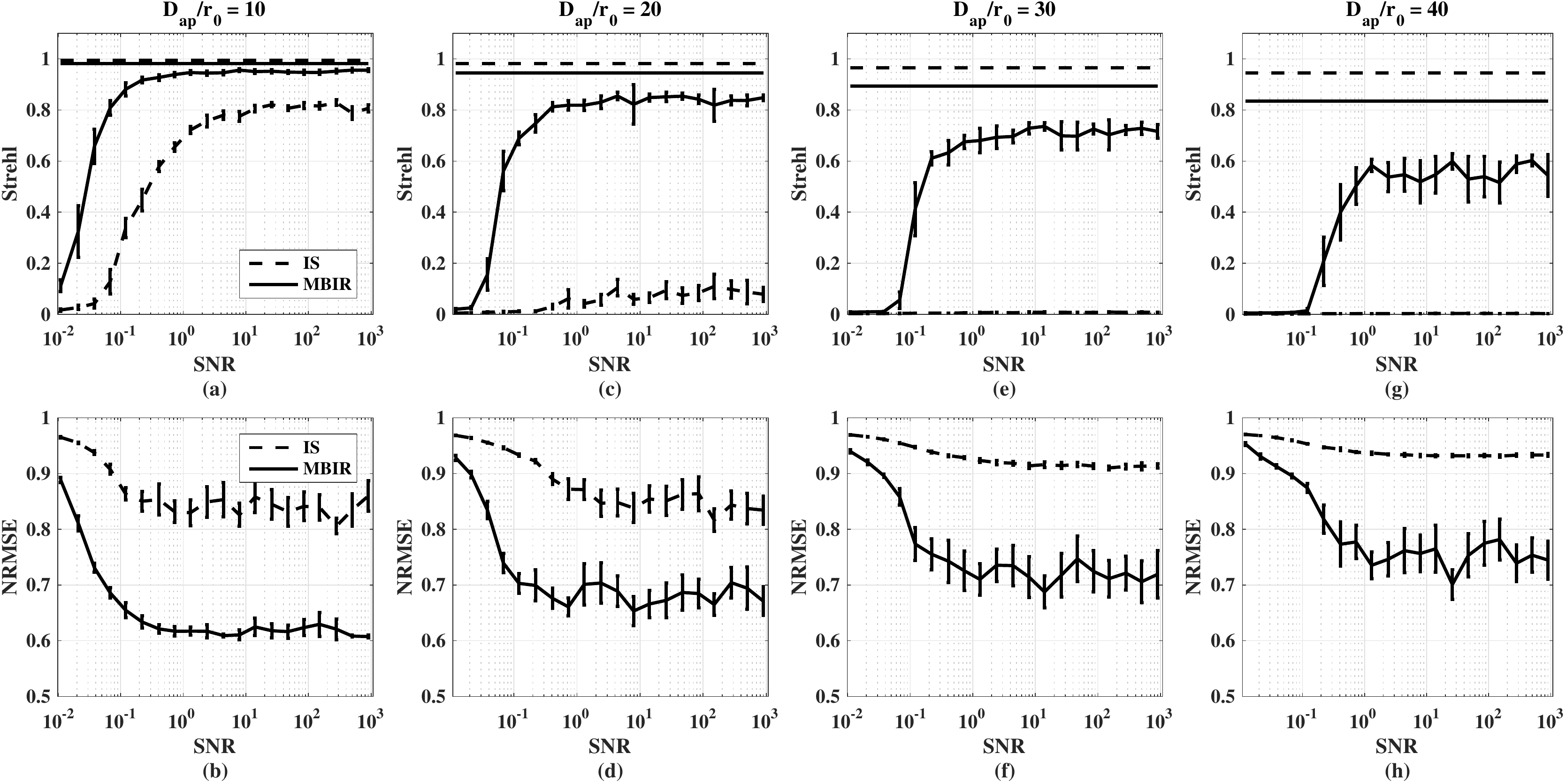}
\caption{Strehl ratio (top row) and reconstruction error (bottom row) vs SNR for $D_{ap}/r_0$ of 10 (a-b),  20 (c-d), 30 (e-f), and 40 (g-h).  The dashed curves shows the IS results and the solid curves show the MBIR results.  The horizontal lines in the top row show the corresponding Strehl ratio limit given by Eq.~(\ref{eq:strehl_lim}).}
\label{fig:results1}
\end{figure*}

Figure~\ref{fig:results1} shows the resulting Strehl ratios and NRMSE for each algorithm as a function of SNR and $D_{ap}/r_0$.  The curves in Fig.~\ref{fig:results1} show the average results for all 18 i.i.d.\ realizations along with error bars which span the standard deviation.  We also plot the Strehl ratio limits given by Eq.~(\ref{eq:strehl_lim}) where we used $d=D_{ap}/256$ for the IS algorithm and $d=D_{ap}/128$ for the MBIR algorithm.

The results show that the IS algorithm works in a limited range of conditions for single-shot data.  For $D_{ap}/r_0=10$, IS peaks at a Strehl ratio of $0.8$ when the SNR exceeds $10$.  For lower SNRs or stronger turbulence, the IS algorithm's performance tapers off quickly.  Conversely, the proposed MBIR algorithm is able to obtain Strehl ratios much higher than IS, even for low SNRs and strong turbulence.  In the $D_{ap}/r_0=10$ case, MBIR reaches the IS Strehl limit of 0.8 with about $92\%$ less SNR.

\section{Conclusion}
In this paper, we presented an inverse, model-based approach to estimating phase errors and reconstructing images from digital holography (DH) data.  We designed the algorithm for cases where only a single data realization is available and the phase errors are isoplanatic.  Rather than estimating the spatially uncorrelated reflection coefficient, $g$, we estimate the highly-correlated reflectance function, $r$.  This allows us to better constrain the underdetermined system and produce speckle-reduced images and accurate phase-error estimates with less data. Using first principals, we derived a discrete forward model for use in the MAP cost function.  To obtain a more-tractable surrogate function, we used the EM algorithm.  Additionally, we introduced a GMRF prior model for the phase error function modeled on a lower resolution grid and presented optimization schemes for the joint estimates.

We compared the proposed algorithm to a leading image sharpness approach over a range of conditions.  The results showed that the MBIR algorithm produced phase estimates with higher Strehl ratios and lower image distortion than the IS technique.  For cases of high noise and large phase errors, IS was not effective, while MBIR was still able to produce accurate estimates.  In conclusion, we showed that the proposed algorithm is an effective alternative to IS algorithms for estimating phase errors from single shot DH data and reconstructing images.

\appendix

\section*{Appendix A: Forward Model Derivation}
In this section we derive the linear forward model expressed by Eq.~(\ref{eq:Data}).  Following the scenario depicted in Fig.~\ref{fig:DH_Fig}, we derive a continuous model for the data, then a discrete representation.

Assume that we use a monochromatic plane wave to illuminate an object which has a complex-valued reflection coefficient, $\tilde{g}(\xi,\eta)$.  We denote continuous functions using tildes.  The return field passing through the entrance pupil can be represented using the Fresnel diffraction integral~\cite{Goodman05} given by
\begin{equation}
\begin{aligned}
	\label{eq:Rx2}
	\tilde{U}_s(x,y)  =  &~ \alpha  ~\tilde{a}(x,y)~e^{j \tilde{\psi}(x,y)}  e^{j \frac{k}{2z} (x^2+y^2)}  \ \\
	&~  \iint  \left\{ \tilde{g}(\xi,\eta) e^{j \frac{k}{2z} (\xi^2+\eta^2)} \right\} e^{-j \frac{k}{z} (x \xi+y \eta)}d \xi d \eta , 
\end{aligned}
\end{equation}
where $\alpha$ is a complex constant, $\tilde{a}(x,y)$ represents the circular transmission function for the system's entrance pupil, $\tilde{\psi}(x,y)$ is the atmospheric phase-error function assumed to be concentrated in a single layer in the entrance-pupil plane (i.e. an isoplanatic phase error), $k=2\pi/\lambda$ is the wave number, and $\lambda$ is the wavelength.

To simplify notation, we define the function
\begin{equation}
\begin{aligned}
	\label{eq:Rx3}
	\tilde{f}(\xi,\eta)  = \tilde{g}(\xi,\eta) e^{j \frac{k}{2z} (\xi^2+\eta^2)} ,
\end{aligned}
\end{equation}
and its corresponding continuous space Fourier transform (CSFT) given by
\begin{equation}
\begin{aligned}
	\label{eq:F_CFT}
\tilde{F}(x,y)=  \iint_{-\infty}^{\infty} \tilde{f}(\xi,\eta) e^{-j2 \pi \left( x\xi + y\eta \right)} \ .
\end{aligned}
\end{equation}
We can also combine the pupil-plane quadratic phase function of Eq.~(\ref{eq:Rx2}) with the atmospheric phase errors to get a single unknown phase term given by
\begin{equation}
\begin{aligned}
	\label{eq:Rx4}
	\tilde{\phi}(x,y)  =  \tilde{\psi}(x,y)+\frac{k}{2z} (x^2+y^2) \ .
\end{aligned}
\end{equation}

Given the definitions in Eqs.~(\ref{eq:Rx3}) through~(\ref{eq:Rx4}), the return signal of Eq.~(\ref{eq:Rx2}) can be written as
\begin{equation}
\begin{aligned}
	\label{eq:Rx5}
	\tilde{U}_s(x,y) =  \alpha~\tilde{a}(x,y) ~e^{j \tilde{\phi}(x,y)} ~\tilde{F}\left( \frac{x}{\lambda z}, \frac{y}{\lambda z} \right) \ . 
\end{aligned}
\end{equation}

Optical heterodyne detection is performed by mixing the returned signal with a local reference beam given by
\begin{equation}
	\label{eq:Lo}
	\tilde{U}_r(x,y) = R_0  e^{j 2 \pi (k_x x + k_y y)},
\end{equation}
where $R_0$ is the amplitude and $k_x$, $k_y$ are factors which control the spatial-frequency modulation.  We combine the signal and reference onto the detector which measures the intensity given by
\begin{equation}
\begin{aligned}
	\label{eq:Intensity}
	\tilde{I}(x,y)  = &~| \tilde{U}_r(x,y) + \tilde{U}_s(x,y)|^2 \\
	 = & ~| \tilde{U}_r(x,y) |^2 + | \tilde{U}_s(x,y)|^2 \\
	&~+  \tilde{U}_r(x,y)\tilde{U}_s^*(x,y) + \tilde{U}_r^*(x,y)\tilde{U}_s(x,y) \\
	= & ~R_0^2 + \alpha^2 \tilde{a}(x,y)^2|\tilde{F}\left( \frac{x}{\lambda z}, \frac{y}{\lambda z} \right)|^2 \\
	& + \alpha~\tilde{a}(x,y) R_0 e^{-j \tilde{\phi}(x,y)} \tilde{F}^*\left( \frac{x}{\lambda z}, \frac{y}{\lambda z} \right)e^{j 2 \pi (k_x x + k_y y)}\\
	&  + \alpha ~\tilde{a}(x,y) R_0 e^{j \tilde{\phi}(x,y)} \tilde{F}\left( \frac{x}{\lambda z}, \frac{y}{\lambda z} \right)e^{-j 2 \pi (k_x x + k_y y)} ,
\end{aligned}
\end{equation}
where $^*$ indicates the complex conjugate.  We assume the system is shot-noise limited with noise $\tilde{w}(x,y)$, driven by the power of the reference beam~\cite{protopopov2009}.  Thus, we model $\tilde{w}(x,y)$ as additive, zero-mean, white Gaussian noise~\cite{Lucke02}.  After detection we demodulate the detector output to remove the spatial-frequency offset of the reference, low-pass filter, and decimate the data.  This isolates the last term of Eq.~(\ref{eq:Intensity}) which gives us our signal of interest \footnote{In practice, the output of the detection process is only proportional to the right-hand side of Eq.~(\ref{eq:cont_sig}) by some unknown constant.},
\begin{equation}
	\label{eq:cont_sig}
	\tilde{y}(x,y) = ~\tilde{a}(x,y)~e^{j \tilde{\phi}(x,y)} \tilde{F}\left( \frac{x}{\lambda z}, \frac{y}{\lambda z} \right) + \tilde{w}(x,y)\ .
\end{equation}

Equation~(\ref{eq:cont_sig}) is continuous in both the $x-y$ and $\xi - \eta$ coordinate systems.  However, we wish to represent the signal as discrete measurements, $y(m,n)$, generated from a discrete-space signal, $f(p,q)$.  We start by representing the discrete field in the object plane as
\begin{equation}
	\label{eq:dis_obj}
	f(p,q) = \tilde{f}(\xi,\eta)\Bigr|_{\substack{\xi=pT_{\xi}~\\ \eta = q T_{\eta}}},
\end{equation}
where $T_{\xi}$ and $T_{\eta}$ are the spatial-sampling periods in the object plane. Furthermore, if we sample the signal with a focal-plane array and ignore the blurring effects of finite-sized pixels, then the discrete measurements can be represented as
\begin{equation}
\begin{aligned}
	\label{eq:dis_data}
	y(m,n) = & \tilde{y}(x,y) \Bigr|_{\substack{y=mT_y~\\ x = nT_x}}, \\
\end{aligned}
\end{equation}
where $T_x$ and $T_y$ are the spatial-sampling periods in the measurement plane.  Combining Eqs.~(\ref{eq:cont_sig})-(\ref{eq:dis_data}), we get 
\begin{equation}
\begin{aligned}
	\label{eq:discrete1}
	y(m,n) = ~a(m,n)~e^{j \phi(m,n)} \sum_{p,q}f(p,q) D_{m,n;p,q}+ w(m,n) ,
\end{aligned}
\end{equation}
where $a$, $\phi$, and $w$ are discrete versions of $\tilde{a}$, $\tilde{\phi}$, and $\tilde{w}$, respectively, and 
\begin{equation}
\begin{aligned}
	\label{eq:discrete2}
	D_{m,n;p,q} = \frac{1}{\sqrt{N}} \exp \left\{ -j 2 \pi \left(  \frac{ T_{\xi} T_x }{\lambda z} m p + \frac{ T_{\eta} T_y } {\lambda z} n q \right) \right\} \ .
\end{aligned}
\end{equation}

Equation~(\ref{eq:discrete1}) can represented more compactly using matrix-vector notation as
\begin{equation}
	\label{eq:LA1}
	y =  A f + w ,
\end{equation}
where $w \in \mathbb{C}^{M}$ is the vectorized measurement noise and $f \in \mathbb{C}^{M}$ is the vectorized field in the object plane.  The matrix $A$ can be decomposed as
\begin{equation}
	\label{eq:LA2}
	A =  \mathcal{D} \left( a \right) \mathcal{D}\left( \exp \left\{ j \phi \right\} \right) D  \ .
\end{equation}
Here $\mathcal{D} \left( \cdot \right)$ denotes an operator that produces a diagonal matrix from its vector argument, $a \in \mathbb{R}^{N}$ is the vectorized entrance-pupil transmission function, and $\phi \in \mathbb{C}^{N}$ is the vectorized phase error function.  The matrix $D$ is defined in Eq.~(\ref{eq:discrete2}).  

Since the sum in Eq.~(\ref{eq:discrete1}) represents the forward propagation of the the field $f$, we have scaled $D$ by $1/\sqrt{N}$ so that $D^HD = I$.  This ensures we conserve energy when propagating between the object and entrance-pupil planes.  Furthermore, we choose our reconstruction parameters such that $T_{\xi} T_x/ \lambda z = N_x$ and $T_{\eta} T_y/ \lambda z = N_y$, where $N_x$ and $N_y$ are the grid sizes in the $x$ and $y$ dimensions, respectively.  Thus, $D$ is exactly a Discrete Fourier Transform (DFT) kernel and can be efficiently implemented using a Fast Fourier Transform (FFT).

\section*{Appendix B: Derivation of the EM Surrogate Function}


In this section, we derive the EM surrogate for the MAP cost function.  We start by writing Eq.~(\ref{eq:Surrogate1}) as
\begin{equation}
\begin{aligned}
	\label{eq:MAPCost2b}
	Q(r,\phi;r',\phi')  = & - \text{E} \left[ \log p(y,f~ | r,\phi) | Y =y, r', \phi' \right] \\
	                            & ~~~- \log p \left( r \right) - \log p \left( \phi \right),\\ 
	                          = & - \text{E} \left[ \log p(y|f,\phi) + \log p(f | r) ~| Y =y, r', \phi' \right] \\
	                             & ~~~- \log p \left( r \right) - \log p \left( \phi \right), \\
	 \end{aligned}
\end{equation}
where we have used Bayes' theorem inside the expectation and the fact that $p(y|f,r,\phi)=p(y|f,\phi)$.  Next, we substitute in the forward and prior models specified in sections~\ref{Estimation_Framework} and~\ref{Prior_Models}.   This gives
\begin{equation}
\begin{aligned}
	\label{eq:MAPCost3}
	Q(r,\phi;r',\phi') = & ~ \text{E} \left[ \frac{1}{\sigma_w^2} ||y-A_{\phi} f||^2~| Y =y, r', \phi' \right] + \log | \mathcal{D}( r)| \\
	 &  ~~~+ \sum_{i=1}^{N} \frac{1}{r_i}~\text{E} \left[  | f_i |^2~| Y =y, r', \phi' \right] + \sum_{\left\{ i,j \right\} \in \mathcal{P} } b_{i,j} \rho_r \left( \frac{\Delta_r}{\sigma_r}\right)\\
	 & ~~~ + \sum_{\left\{ i,j \right\} \in \mathcal{P} } b_{i,j} \rho_{\bar{\phi}} \left( \frac{\Delta_{\bar{\phi}}}{\sigma_{\bar{\phi}}}\right) + \kappa,
 \end{aligned}
\end{equation}
where $A_{\phi}$ indicates the matrix $A$ is dependent on $\phi$ and the variable $\kappa$ is a constant with respect to $r$ and $\phi$.     

To evaluate the expectations in Eq.~(\ref{eq:MAPCost3}), we must specify the conditional posterior distribution of $f$.  Using Bayes' theorem, 
\begin{equation}
	\label{eq:PostCondDist1}
	\begin{aligned}
	p(f|y,r,\phi)	& = \frac{p(y|f,r,\phi) p(f|r)}{p(y|r,\phi)},\\
	& = \frac{1}{z} \exp \left\{  -\frac{1}{\sigma_w^2 } ||y-A_{\phi}f||^2- f^H \mathcal{D}( r)^{-1}f\right\},
	\end{aligned}
\end{equation}
where $z$ is the partition function which absorbs any exponential terms that are constant with respect to $f$.  By completing the square, we can show that the posterior distribution is a complex Gaussian with mean
\begin{equation}
	\label{eq:mu_g}
	\mu=C  \frac{1}{\sigma_w^2 } A_{\phi'}^H y,
\end{equation}
and covariance
\begin{equation}
\begin{aligned}
	\label{eq:C_g}
	C & = \left[\frac{1}{\sigma_w^2 } A_{\phi'}^HA_{\phi'}+  \mathcal{D}( r)^{-1}\right]^{-1} \ .
\end{aligned}
\end{equation}

Using the posterior distribution specified by Eq.~(\ref{eq:PostCondDist1}), we can evaluate the expectations in Eq.~(\ref{eq:MAPCost3}) to get the final form of our EM surrogate function given by
\begin{equation}
\begin{aligned}
	\label{eq:MAPCost4}
	Q(r,\phi;r',\phi') = &  - \frac{1}{\sigma_w^2} 2 \text{Re} \left\{ y^HA_{\phi} \mu \right\}  + \log | \mathcal{D}( r)| \\
	& ~~~+ \sum_{i=1}^{N} \frac{1}{r_i}~\left( C_{i,i} + | \mu_i|^2\right)  + \sum_{\left\{ i,j \right\} \in \mathcal{P} } b_{i,j} \rho_r \left( \frac{\Delta_r}{\sigma_r}\right)\\
	& ~~~ + \sum_{\left\{ i,j \right\} \in \mathcal{P} } b_{i,j} \rho_{\bar{\phi}} \left( \frac{\Delta_{\bar{\phi}}}{\sigma_{\bar{\phi}}}\right) + \kappa , \\
	 \end{aligned}
\end{equation}
where $\mu_i$ is the $i^{th}$ element of the posterior mean and $C_{i,i}$ is the $i^{th}$ diagonal element of the posterior covariance.

\section*{Acknowledgment}
We would like to thank Dr. Samuel T. Thurman for his helpful discussions and support of this work.  In addition, we would also like to thank the Maui High Performance Computing Center (MHPCC) for their assistance and resources which allowed us to test the proposed algorithm over a wide range of conditions.

\bibliography{all}

\ifthenelse{\equal{\journalref}{ol}}{%
\clearpage
\bibliographyfullrefs{sample}
}{}
 

\ifthenelse{\equal{\journalref}{aop}}{%
\section*{Author Biographies}
\begingroup
\setlength\intextsep{0pt}
\begin{minipage}[t][6.3cm][t]{1.0\textwidth} 
  \begin{wrapfigure}{L}{0.25\textwidth}
    \includegraphics[width=0.25\textwidth]{john_smith.eps}
  \end{wrapfigure}
  \noindent
  {\bfseries John Smith} received his BSc (Mathematics) in 2000 from The University of Maryland. His research interests include lasers and optics.
\end{minipage}
\begin{minipage}{1.0\textwidth}
  \begin{wrapfigure}{L}{0.25\textwidth}
    \includegraphics[width=0.25\textwidth]{alice_smith.eps}
  \end{wrapfigure}
  \noindent
  {\bfseries Alice Smith} also received her BSc (Mathematics) in 2000 from The University of Maryland. Her research interests also include lasers and optics.
\end{minipage}
\endgroup
}{}

\end{document}